\documentclass[12pt]{article}

\usepackage[x11names]{xcolor}
\usepackage{tikz}
\usepackage{cite}
\usepackage{comment}
\usepackage{hyperref}
\usetikzlibrary{decorations.pathreplacing,decorations.markings}
\usepgflibrary{arrows}
\usepgflibrary[arrows]
\usetikzlibrary{arrows}
\usetikzlibrary[arrows]
\usetikzlibrary{positioning}
\usepackage{graphicx}
\usepackage{amsmath}
\usepackage{amssymb}
\usepackage{subsupscripts}
\usepackage{braket}
\usepackage{amsthm}
\usepackage{xcolor}

\makeatletter

\@addtoreset{equation}{section}
\makeatother
\newcommand{\ba}{\begin{eqnarray}}
\newcommand{\ea}{\end{eqnarray}}

\newcommand{\bpsi}{\bar\psi}

\usetikzlibrary{decorations.pathreplacing,decorations.markings}
\tikzset{
  on each segment/.style={
    decorate,
    decoration={
      show path construction,
      moveto code={},
      lineto code={
        \path [#1]
        (\tikzinputsegmentfirst) -- (\tikzinputsegmentlast);
      },
      curveto code={
        \path [#1] (\tikzinputsegmentfirst)
        .. controls
        (\tikzinputsegmentsupporta) and (\tikzinputsegmentsupportb)
        ..
        (\tikzinputsegmentlast);
      },
      closepath code={
        \path [#1]
        (\tikzinputsegmentfirst) -- (\tikzinputsegmentlast);
      },
    },
  },
  mid arrow/.style={postaction={decorate,decoration={
        markings,
        mark=at position .5 with {\arrow[#1]{stealth}}
      }}},
}

\newcommand{\dket}[1]{\ket{#1}\!\rangle}
\newcommand{\dbra}[1]{\langle\!\bra{#1}}

 \def\d{\delta}

 \def\p{\partial}
 
 \def\a{\alpha}
 
 \def\g{\gamma}
 \def\d{\delta}
 \def\e{\varepsilon}

 \def\k{\kappa}
 \def\l{\lambda}

 \def\s{\sigma}
 \def\t{\tau}

 \def\G{\Gamma}
 \def\D{\Delta}

\def\CF{{\mathcal{F}}}

\def\CN{{\mathcal{N}}}

\def\CS{{\mathcal{S}}}
\def\CT{{\mathcal{T}}}

\def\CV{{\mathcal{V}}}

\def\CZ{{\mathcal{Z}}}
\def\la{\left\langle}
\def\ra{\right\rangle}

\def\hf{\dfrac{1}{2}}
\def\bc{{\bar{c}}}

\def\implies{\quad\Rightarrow\quad}

\def\tW{{\tilde{W}}}

\def\CS{\mathcal{S}}

\def\vac{\emptyset}

\def\mZ{\mathbb{Z}}
\def\mR{\mathbb{R}}
\def\mC{\mathbb{C}}

\def\bW{{\bar W}}
\def\tW{\tilde{W}}
\def\gl{\mathfrak{gl}}

\def\glinf{\widehat{\mathfrak{gl}}(\infty)}

\def\Abox{{\tikz[scale=0.007cm] \draw (0,0) rectangle (1,1);}}
\def\sAbox{{\tikz[scale=0.005cm] \draw (0,0) rectangle (1,1);}}

\def\sAcube{\tikz[scale=0.005cm] \draw (0,0,0) rectangle (1,1,0) (0,1,0) -- (0,1,-1) -- (1,1,-1) --(1,1,0) (1,0,0) -- (1,0,-1) -- (1,1,-1);}
\def\Winf{W_{1+\infty}}

\def\ad{\text{ad}}
\def\bst{\boldsymbol{t}}

\begin{document}
\begin{titlepage}
\vspace*{-2cm}
\begin{flushright}
CQUEST-2021-0658
\end{flushright}
\vspace*{1cm}
\vskip 12mm

\begin{center}
    {\Huge Intertwining operator and integrable hierarchies\\
     \vskip .5cm
    from topological strings}
    \vskip 2cm
    {\Large Jean-Emile Bourgine}\\
    \vskip 2cm
    {\it Center for Quantum Spacetime (CQUeST)}\\
    {\it Sogang University}\\
    {\it Seoul, 121-742, South Korea}
    \vskip 1cm
    {\it Korea Institute for Advanced Studies (KIAS)}\\
    {\it Quantum Universe Center (QUC)}\\
    {\it 85 Hoegiro, Dongdaemun-gu, Seoul, South Korea}
    \vskip 1cm
    \texttt{bourgine@kias.re.kr}
\end{center}
\vfill
\begin{abstract}
In \cite{Nakatsu2007}, Nakatsu and Takasaki have shown that the melting crystal model behind the topological strings vertex provides a tau-function of the KP hierarchy after an appropriate time deformation. We revisit their derivation with a focus on the underlying quantum $\Winf$ symmetry. Specifically, we point out the role played by automorphisms and the connection with the intertwiner - or vertex operator - of the algebra. This algebraic perspective allows us to extend part of their derivation to the refined melting crystal model, lifting the algebra to the quantum toroidal algebra of $\gl(1)$ (also called Ding-Iohara-Miki algebra). In this way, we take a first step toward the definition of deformed hierarchies associated to A-model refined topological strings.
\end{abstract}
\vfill
\end{titlepage}

\setcounter{footnote}{0}

\newpage

\section{Introduction}
The infinite Lie algebra $\glinf$ and its fermionic representation plays a key role in the construction of solutions for the Kadomtsev-Petviashvili (KP) hierarchy \cite{Sato1983}. This role was emphasized by the Kyoto school \cite{Date1981a}, and led to the introduction of several reductions of the KP hierarchy associated to subalgebras of $\glinf$. This algebraic approach was further formalized by Kac and Wakimoto, and extended to associate a hierarchy of differential equations to any Kac-Moody algebra \cite{Kac1989,Kac1990}. These works illustrate the efficiency of the algebraic approach for integrable hierarchies.\footnote{For a brief introduction, see the excellent reviews \cite{Jimbo1983,Alexandrov2012} or the reference book \cite{Kac1990}.}

Another indisputable success of the algebraic approach is in the study of 2D Conformal Field Theories (CFT) where most physical quantities are determined by the symmetry algebras \cite{Belavin1984}. This formidable success led Frenkel and Reshetikhin to propose an extension of the vertex operator technique employed in 2D CFT to some quantum integrable systems using their quantum affine algebra of symmetries. Indeed, algebraically, a vertex operator is simply an intertwiner between a Fock (or level one) representation $\CF$ of the algebra and its tensor product $\CV\otimes\CF$ with a level zero representation $\CV$. For the original vertex operator of a free boson, the algebra is a Heisenberg algebra, it becomes an affine Lie algebra in Wess-Zumino-Witten models. In the case of quantum affine algebras, the Frenkel-Jing \cite{Frenkel1988} construction provides the level one representation while the level zero representation is the usual highest weight representation acting on the quantum spins. This vertex operator technique has been applied, for instance, to the diagonalization of the infinite XXZ spin chain Hamiltonian in \cite{Davies1992}. To avoid the confusion with exponentials of free fields, the vertex operator is often called \textit{intertwining operator}, or simply \textit{intertwiner}, in this context.

In \cite{AFS}, Awata, Feigin and Shiraishi (AFS) applied the vertex operator technique for a different algebra called quantum toroidal $\gl(1)$ (or Ding-Iohara-Miki) algebra \cite{Ding1997,Miki2007}. They have shown that, quite remarkably, the matrix elements of the intertwiner reproduce the refined vertex of the topological string theory \cite{AKMV,Iqbal2007,Awata2008}. This observation led to a number of important results in this field. For example, we can mention the extension of the topological vertex technique to various theories \cite{BFMZ,Awata2017,Foda2018,Zenkevich2018,Bourgine2018,Kimura:2019gon,Bourgine:2019phm,Zenkevich2020} and observables \cite{BMZ,Bourgine2016,Bourgine2017b}, the derivation of proofs for the q-deformed AGT correspondence \cite{Feigin2010,Awata2011,Fukuda2019}, or the description of the fiber-base duality \cite{Awata2009a,Bourgine2018a,SWM,Fukuda2019}.

In this paper, we investigate the role played by quantum algebras in the well-known relation between self-dual topological strings and integrable hierarchies \cite{Aganagic2003}. Refined topological strings depend on two parameters $(q,t^{-1})$, they are identified with the parameters $(q_1,q_2)$ of the quantum toroidal $\gl(1)$ algebra. In the self-dual limit $t\to q$ (or $q_1q_2\to 1$), the original formulation of topological strings is recovered, the parameter $q$ is identified with the exponentiated string coupling constant $e^{g_\text{str}}$ and the algebra reduces to the quantum $W_{1+\infty}$ algebra \cite{KR,AFMO}. The latter is equivalent to the $\glinf$ algebra mentioned earlier upon a linear transformation of the generators \cite{Lebedev1992,Hoppe1993,Bourgine2021}, it is thus naturally expected to be involved in the relation with integrable hierarchies, which was indeed observed in \cite{Aganagic2003}.

In the melting crystal picture developed by Okounkov, Reshetikhin and Vafa \cite{ORV}, the self-dual topological vertex is interpreted as the generating function of plane partitions weighted by the number of boxes. In \cite{Nakatsu2007}, Nakatsu and Takasaki have shown that this generating function, once properly deformed by extra time dependencies, provides a tau function for the KP hierarchy. Although the role played by the quantum $\Winf$ algebra in this setup has already been described in \cite{Nakatsu2007}, we revisit here their derivation with an extra emphasis on certain algebraic structures. Specifically, we underline the role of two particular objects, the first one being the AFS intertwining operator introduced previously, and the second one is an operator associated to the framing factors of the topological vertex that we call \textit{framing operator} for short. Algebraically, it realizes the action of a certain automorphism $\CT$ in the Fock module of the algebra. As we shall see, these two objects are deeply related to the $SL(2,\mZ)$ subgroup of automorphisms. Both are essential ingredients underlying the Nakatsu-Takasaki derivation.

The main motivation for our study is the search for a deformation of integrable hierarchies that would correspond to the refinement of topological strings in their A-model formulation. Naively, it would seem difficult as the fermionic structure disappears from the Fock representation, and only a bosonic formulation subsists. Yet, the presence of the quantum toroidal algebra hints for the preservation of an integrable structure. It brings us to the second part of our paper in which we attempt to refine the Nakatsu-Takasaki derivation. The definition and properties of the two previous algebraic objects are easily extended to the quantum toroidal case, and we manage to write the generating function in the canonical form 
\begin{equation}\label{canonical}
\bra{\vac}e^{-\sum_k\frac1k\t_kJ_k}g\ket{\vac},
\end{equation}
where $J_k$ are the modes of a Heisenberg algebra, $\ket{\vac}$ the vacuum of the Fock space and $\t_k$ the (rescaled) times of the hierarchy. Unfortunately, we were unable to show that $g$ is a group-like element of $GL(\infty)$, that would imply that the generating function is a tau function of an integrable hierarchy. And, in fact, it cannot be group-like as a perturbative analysis indicates that the Hirota equation is no longer satisfied in the refined case. Nevertheless, we are confident that this result is bringing us closer to the definition of refined hierarchies.

The organization of the paper is very straightforward: the first part deals with self-dual topological strings and the quantum $\Winf$ algebra while the second part presents the refined case corresponding to the quantum toroidal algebra. The appendix contains the proofs for several identities used in the main text.

\section{Integrable structure of the melting crystal model}
\subsection{Quantum $\Winf$ algebra}
\subsubsection{Definition}
The quantum $\Winf$ algebra can be presented in many ways and found under different names, e.g. quantum torus algebra in \cite{Nakatsu2007}, trigonometric Sin-Lie algebra in \cite{FFZ1989,Lebedev1992},... Here, we use a presentation in terms of generators $W_{m,n}$ with integer indices, and two central elements $(c_1,c_2)$, that satisfy the commutation relations
\begin{equation}\label{def_qW}
[W_{m,n},W_{m',n'}]=(q^{m'n}-q^{mn'})\left(W_{m+m',n+n'}+c_1\dfrac{\d_{m+m'}}{1-q^{n+n'}}-c_2\dfrac{\d_{n+n'}}{1-q^{m+m'}}\right).
\end{equation}
In fact, this algebra is a central extension by the element $c_2$ of the quantum algebra used in \cite{Nakatsu2007} (up to a rescaling of the generators). In application to integrable hierarchies, we consider mostly the representation of levels $(1,0)$, in which case the two algebras coincide. However, the presence of a second central element is important in the description of automorphisms. A second motivation for the introduction of two central charges is their interpretation in the algebraic engineering of topological strings amplitudes. Indeed, in this context, the two levels are identified with the label of the degenerating cycles in the toric diagram of the Calabi-Yau \cite{Mironov2016,Awata2016a}. They also correspond to the charge of the branes in the $(p,q)$-brane web construction \cite{Aharony1997,Aharony1997a}.

The algebra has the group of automorphisms $GL(2,\mZ)$ \cite{Hoppe1993}, but we will be only interested in the $SL(2,\mZ)$ subgroup generated by the transformations $\CS$ and $\CT$. These transformations act as follows on the generators,
\begin{align}\label{transfo_ST}
\begin{split}
\CS:\ & W_{m,n}\to q^{-(m+1)n}W_{-n,m},\quad (c_1,c_2)\to (c_2,-c_1),\\
\CT:\ &W_{m,n}\to q^{-n^2/2}\left(W_{m-n,n}+c_1\dfrac{\d_{m-n}-q^{n^2/2}\d_{m,0}}{1-q^n}(1-\d_{n,0})-c_1\dfrac{\d_{n,0}}{1-q^{m}}(1-\d_{m,0})\right),\\
&(c_1,c_2)\to (c_1,-c_1+c_2).
\end{split}
\end{align}

\subsubsection{Dirac module}
The quantum $\Winf$ algebra has a representation of levels $(1,0)$ on the Fock space $\CF$ of a Dirac fermion. We refer the reader to \cite{Bourgine2021} for a recent review of this representation. The Fock space is built from a vacuum state $\ket{\vac}$ annihilated by the positive modes of the fermionic fields\footnote{There are different conventions for the fermionic modes and we use here half-integer indices. To recover the convention employed in \cite{Nakatsu2007}, one should simply replace $\psi_r\to \psi_{r+1/2}^\ast$ and $\bpsi_r\to \psi_{r-1/2}$.}
\begin{equation}\label{fermions}
\psi(z)=\sum_{r\in\mathbb{Z}+1/2}z^{-r-1/2}\psi_r,\quad \bpsi(z)=\sum_{r\in\mathbb{Z}+1/2}z^{-r-1/2}\bpsi_r,\quad \{\psi_r,\bpsi_s\}=\d_{r+s}.
\end{equation}
Due to the bosonization formulas $\bar\psi(z)=:e^{\phi(z)}:$, $\psi(z)=:e^{-\phi(z)}:$, and $:\bpsi(z)\psi(z):=\p\phi(z)$ involving the bosonic field
\begin{equation}\label{exp_phi}
\phi(z)=Q+J_0\log z-\sum_{k\in\mZ^\times}\dfrac1kz^{-k}J_k,\quad [J_k,J_l]=k\d_{k+l},\quad [Q,J_0]=1,
\end{equation}
the Fock space $\CF$ has an alternative construction obtained by the action of the negative modes $J_{-k}$ on the vacuum state $\ket{\vac}$. The vacuum is annihilated by positive modes $J_k$, and the normal ordering naturally consists in moving these modes to the right. The Fock space $\CF$ can be decomposed according to the values of the zero mode $J_0$ into $\CF=\bigoplus_{k\in\mZ} \CF_k$.
Since the representation of the generators $W_{m,n}$ is neutral, we can restrict ourselves to the subspace $\CF_0$ of zero charge. Because of this restriction, we will only be able to discuss the KP hierarchy (not mKP), but it makes the question of refinement simpler.

\paragraph{Symmetric polynomials} The vector space $\CF_0$ is isomorphic to the space of symmetric polynomials with infinitely many variables. The isomorphism sends the vacuum $\ket{\vac}$ to the trivial polynomial $1$, and maps the PBW basis to products of elementary powers sums $p_k(x)=\sum_kx_i^k$,
\begin{equation}
(J_{-\l_1})^{k_1}\cdots (J_{-\l_n})^{k_n}\ket{\vac} \leftrightarrow p_{\l_1}(x)^{k_1}\cdots p_{\l_n}(x)^{k_n}.
\end{equation} 
Thus, the negative modes $J_{-k}$ act as a multiplication by $p_k$ while positive modes $J_k$ act as $k\p/\p p_k$ on symmetric polynomials. This isomorphism is useful in order to define the Schur basis of states $\ket{\l}$, labeled by a partition $\l$, and obtained as the inverse image of the Schur polynomials $s_\l(x)$ (they form a basis of the ring of symmetric polynomials). These Schur states coincide with the fermionic PBW basis of $\CF_0$. In the same way, we define the Macdonald basis with states $\ket{P_\l}$ obtained as the inverse image of the Macdonald polynomials $P_\l(x)$. We also define the dual states $\bra{\l}$ and $\bra{P_\l}$ with the scalar product
\begin{equation}
\bra{\l}\!\!\ket{\mu}=\d_{\l,\mu},\quad \bra{P_\l}\!\!\ket{P_\mu}=\la P_\l,P_\l\ra_{q,t}\d_{\l,\mu},
\end{equation} 
where $\la P_\l,P_\l\ra_{q,t}$ is the norm square of Macdonald scalar product \cite{Macdonald}. Explicitly, 
\begin{equation}\label{norm_Mac}
\la P_\l,P_\l\ra_{q,t}=\prod_{\sAbox\in\l}\dfrac{1-q^{a(\sAbox)+1}t^{l(\sAbox)}}{1-q^{a(\sAbox)}t^{l(\sAbox)+1}},
\end{equation} 
where $a(\Abox)=\l_i-j$ and $l(\Abox)=\l'_j-i$ are the arm and leg length of the box $(i,j)$ in the Young diagram representing the partition $\l=(\l_1,\l_2,\cdots)$. We denote $\l'$ the partition corresponding to the transposed Young diagram of $\l$.

\paragraph{Action of the algebra} We denote $\rho_{u,v}^{(D)}$ the representation of the quantum $\Winf$ algebra of levels $(1,0)$ and weights $u,v\in\mC^\times$ on the Fock space $\CF_0$. To get rid of the weights dependence, we sometimes use the notation $\bW_{m,n}=\rho_{1,1}^{(D)}(W_{m,n})$ so that the general representation reads
\begin{equation}
\rho_{u,v}^{(D)}(W_{m,n})=u^nv^m\bW_{m,n}-\dfrac{1-u^n}{1-q^n}\d_{m,0}(1-\d_{n,0}).
\end{equation} 
The action on $\CF_0$ can be expressed in different ways, the simplest one is in terms of the fermionic fields,
\begin{equation}
\bW_{m,n}=\sum_{r\in\mZ+1/2}q^{-(r+1/2)n}:\bar\psi_{m-r}\psi_r:.
\end{equation} 
Instead, in the bosonic presentation the modes $J_k$ are identified with the generators $\bW_{k,0}$ while the other generators define currents represented as vertex operators. Fortunately, we will not need these expressions here. Finally, the action of the generators $W_{m,n}$ on the Schur basis can be written explicitly. When $m\neq0$, these operators add or remove strips of $|m|$ boxes to the Young diagram labeling the states \cite{Bourgine2021}. On the other hand, the action of the modes $W_{0,n}$ is diagonal and read
\begin{equation}\label{act_W0k}
\rho^{(D)}_{u,v}(W_{0,-k})\ket{\l}=\phi_k(\l,s)\ket{\l}, \quad \phi_k(\l,s)=-(1-q^{k})q^{ks}\sum_{(i,j)\in\l}q^{-(i-j)k}-\dfrac{1-q^{ks}}{1-q^{-k}}.
\end{equation}
The r.h.s. is independent of the weight $v$, but depends on the weight $u=q^{-s}$ through the variable $s$. The eigenvalues $\phi_k(\l,s)$ are coupled to the time evolution of the hierarchy in \cite{Nakatsu2007,Takasaki2018}.

\paragraph{Framing operator} In addition to the generators $\bW_{m,n}$, it is necessary to introduce the following operators acting on the Fock space $\CF_0$,\footnote{These operators can be obtained using the expansion at $q=e^{\e}\to1$ (small strings coupling limit) of
\begin{equation}
\bW_{0,n}=\sum_{r\in\mZ+1/2}q^{-n(r+1/2)}:\bpsi_{-r}\psi_r:=J_0-n\e\left(L_0+\hf J_0\right)+\hf n^2\e^2\left(W_0+L_0+\dfrac14J_0\right)+O(\e^3).
\end{equation}
They correspond to the zero-modes of the currents of spin one, two (a.k.a. Virasoro) and three of the $\Winf$ algebra that appears in the degenerate limit $q\to1$. Note that there is a small mismatch with respect to the definitions given in \cite{Nakatsu2007,Takasaki2013,Takasaki2018} as we introduced shifts by zero modes of lower spin for later convenience. Our operator $W_0$ coincides with the \textit{cut-and-join operator} of ref. \cite{Goulden1997}.}
\begin{equation}\label{def_JLW}
J_0=\sum_{r\in\mZ+1/2}:\bpsi_{-r}\psi_r:,\quad L_0=\sum_{r\in\mZ+1/2}r:\bpsi_{-r}\psi_r:,\quad W_0=\sum_{r\in\mZ+1/2}r^2:\bpsi_{-r}\psi_r:.
\end{equation} 
These operators act diagonally on the Schur states, with the eigenvalues given by
\begin{equation}
J_0\ket{\l}=0,\quad L_0\ket{\l}=|\l|\ \ket{\l},\quad W_0\ket{\l}=\k(\l)\ \ket{\l}.
\end{equation} 
Here $|\l|$ denotes the number of boxes in the Young diagram of $\l$ and $\k(\l)=2\sum_{(i,j)\in\l}(j-i)$. As we shall see, the insertion of the operator $Q^{L_0}$ inside a bosonic correlator is interpreted as the introduction of a K\"ahler parameter $Q$. In the same way, the insertion of the diagonal operator $q^{-nW_0/2}$ introduces a framing factor $q^{-n\k(\l)/2}$. This factor is related to the Chern-Simons factor at level $n$ in the dual 5D $\CN=1$ gauge theory,
\begin{equation}\label{def_CS}
\CZ_{CS}(n)=\prod_{(i,j)\in\l}v^nq^{(i-j)n}=v^{n|\l|}q^{-n\k(\l)/2}.
\end{equation}
The extra dependence in the weight $v$ can be removed using the operator $v^{-nL_0}$, it is usually absorbed in the definition of the instanton counting parameter \cite{Bourgine2017b}.

It has been observed in \cite{Bourgine2017b} that framing factors follow from the action of the automorphism $\CT$ in the algebraic framework. Thus, we expect the operator $W_0$ to be somehow associated to this automorphism. To formulate this relation, it is more convenient to introduce the combination $\tW_0=(W_0+L_0)/2$, and we will call \textit{framing operators} the operators of the form $q^{\a\tW_0}$ for $\a\in\mR$. The shift by $L_0$ is related to the presence of the parameter $v$ in the Chern-Simons factor \ref{def_CS} that will be set to $v=q^{-1/2}$ later on. Then, using the free fermion realization, one can show that
\begin{equation}\label{tW0_action}
q^{\a\tW_0}\bW_{m,n}q^{-\a\tW_0}=q^{\a m^2/2}\bW_{m,n+\a m}.
\end{equation} 
When $\a$ is an integer, the r.h.s. coincides with the representation of the generators $\CS\CT^{\a}\CS^{-1}\cdot W_{m,n}$. In this case, the adjoint action of $q^{\a\tW_0}$ realizes the $\CS$-dual action of the automorphism $\CT^\a$ on the Fock space $\CF_0$. Very remarkably, this formula also makes sense for $\a\in\mR$, somehow extending the automorphism to non-integer values. The relation \ref{tW0_action} has been derived in \cite{Nakatsu2007,Takasaki2013,Takasaki2018} and called a \textit{shift symmetry}, but the connection with the action of the automorphisms appears to be new.

\subsubsection{Intertwiner and melting crystal}
The AFS intertwiner has been introduced directly in the context of the refined topological vertex \cite{AFS}. However, it is relatively easy to perform the self-dual limit $t\to q$ of this object and obtain the formulation relevant to the unrefined case \cite{SWM}. Since the quantum toroidal $\gl(1)$ algebra reduces to quantum $\Winf$ in this limit, we obtain an intertwiner between modules of the latter. The automorphisms $\CS$ and $\CT$ are then used map these modules to the Fock space $\CF_0$, twisting the Dirac representation in the process. As a result, the intertwiner $\Phi$ is defined as the operator $\CF_0\otimes \CF_0\to \CF_0$ solving the following equation\footnote{It may be useful to make a short historical comment about this equation. In 2D CFT, primary operators of conformal dimension $h$ satisfy the equation $[L_n,\phi_h(z)]=(z^{n+1}\p_z+h(n+1)z^n)\phi_h(z)$ with $n\geq1$ under the adjoint action of the Virasoro modes $L_n$. This equation can be written in the form \ref{AFS_lemma}, i.e. $(\rho_0\otimes\rho_1\ \D(L_n))\ \phi_h(z)=\phi_h(z)\rho_1(L_n)$, where $\rho_1$ is a Fock representation and $\rho_0(L_n)=-z^{n+1}\p_z-h(n+1)z^n$ is the level zero representation that describes the action of the conformal symmetry on the coordinates (the coproduct is also co-commutative). In addition, vertex operators of the free boson obey a similar equation with the Heisenberg algebra $[J_n, V_\a(z)]=-\a z^nV_\a(z)$ where the r.h.s. is trivially a representation of level zero. This equation is generalized further to Wess-Zumino-Witten models and affine Lie algebras \cite{DiFrancesco1997}.}
\begin{equation}\label{AFS_lemma}
\rho_{u_{11},v_{11}}^{(D)}(\CT\cdot W)\Phi(v)=\Phi(v)\left(\rho_{u_{01},v_{01}}^{(D)}\circ \CS\otimes \rho_{u_{10},v_{10}}^{(D)}\ \D(W)\right),
\end{equation}
where $W$ denotes any element of the quantum $\Winf$ algebra. The co-algebraic structure of the quantum toroidal algebra trivializes in the self-dual limit and thus the coproduct in the r.h.s. is the co-commutative one, i.e. $\D(W)=W\otimes1+1\otimes W$. The intertwiner depends on a free parameter $v$, and the weights of the representations are required to obey the three constraints $v_{11}=v_{10}$, $u_{01}=qvv_{11}$, and $u_{11}=-q^{1/2}vv_{11}u_{10}$. It is worth mentioning that a dual intertwiner $\Phi^\ast:\CF_0\to\CF_0\otimes \CF_0$ is also introduced in \cite{AFS} even though it will not be needed here.

The first space $\CF_0$ in the tensor product $\CF_0\otimes \CF_0$ corresponds to the vertical module in the language of the toroidal algebra, it is associated to the preferred direction of the topological vertex and it plays the role of the level zero representation for the vertex operator. Even in the self-dual limit, the formalism retains this notion of a preferred direction, and it is useful to introduce a notation to distinguish it. Following our earlier works \cite{Bourgine2016,Bourgine2017b,Bourgine2018a}, we denote with a double ket (e.g. $\dket{\l}$) the vectors of this module. The solution of the AFS equation \ref{AFS_lemma} is nicely expressed by decomposition on the Schur basis of the vertical module, each component corresponding to a vertex operator $\Phi_\l:\CF_0\to\CF_0$, \begin{equation}\label{vert_dec}
\Phi(v)=\sum_\l\dbra{\l}\otimes\Phi_\l(v),\quad \Phi_\l(v)=t_\l(v):\Phi_\vac(v)\prod_{(i,j)\in\l}\eta(vq^{i-j}):.
\end{equation}
In the second equation, the component $\Phi_\l(v)$ has been decomposed into a normalization factor $t_\l$, a vacuum contribution and a dressing by vertex operators $\eta(z)$,
\begin{equation}\label{def_Phi}
\Phi_\vac(v)=:\exp\left(\sum_{k\in\mZ^\times}\dfrac{v^{-k}}{k(1-q^k)}J_k\right):,\quad \eta(z)=:\exp\left(-\sum_{k\in\mZ^\times}\dfrac{z^{-k}}{k}(1-q^{-k})J_k\right):.
\end{equation}
The normalization factor simplifies if we impose the extra constraint $q^{-3/4}u_{11}v_{01}=v_{11}v^{1/2}$ among the weights: it no longer depends on $v$ and simply writes $t_\l=s_{\l'}(q^{-\rho})$ with $\rho=(-1/2,-3/2,\cdots)$.\footnote{The notation $s_{\l'}(q^{-\rho})$ refers to the evaluation of the Schur polynomial $s_{\l'}(x)$ for the variables $(x_1,x_2,\cdots)=(q^{1/2},q^{3/2},...)$. We assume $|q|<1$ for convergence issues.}

\paragraph{Shift symmetries} The key ingredients in the derivation of the tau function from the melting crystal are a set of relations called \textit{shift symmetries} in \cite{Nakatsu2007,Takasaki2013,Takasaki2018}. We have already encountered one of these relations in \ref{tW0_action}. The other relations are of a different nature: they follow from the AFS intertwining equation \ref{AFS_lemma} by projection on the vacuum of the vertical module. In order to show this fact, we introduce the vertical decomposition \ref{vert_dec} inside the equation \ref{AFS_lemma} and project it on the component $\l$,
\begin{equation}\label{AFS_lemma_2}
\rho_{u_{11},v_{11}}^{(D)}(\CT\cdot W)\Phi_\l(v)=\Phi_\l(v)\rho_{u_{10},v_{10}}^{(D)}(W)+\sum_\mu\Phi_\mu(v)\dbra{\mu}\rho_{u_{01},v_{01}}^{(D)}(\CS\cdot W)\dket{\l}.
\end{equation}
Then, we observe that the action of the modes $W_{m,n}$ with $m>0$ on the Schur states $\dket{\l}$ is to remove strips of $m$ boxes to the partition $\l$, and, in particular, these modes annihilate the vacuum $\dket{\vac}$ \cite{Bourgine2021}. Applying this property to the $\CS$-dual modes $\CS\cdot W_{m,n}=q^{-(m+1)n}W_{-n,m}$ in the previous equation, we obtain an exchange relation for the vacuum component $\l=\vac$,
\begin{equation}\label{AFS_proj}
\rho_{u_{11},v_{11}}^{(D)}(\CT\cdot W_{m,n})\Phi_\vac(v)=\Phi_\vac(v)\rho_{u_{10},v_{10}}^{(D)}(W_{m,n}),\quad n<0.
\end{equation}
This exchange relation reproduces the two missing shift symmetries, as will become clear once we review the connection between the intertwiner and the melting crystal formalism.

\paragraph{Melting crystal} In \cite{ORV}, Okounkov, Reshetikhin and Vafa (ORV) discovered an intriguing connection between the topological vertex \cite{AKMV} and plane partitions. They interpreted the vertex as the generating function of plane partitions with fixed asymptotics given by the three Young diagrams $\l,\mu,\nu$ labeling the vertex $C_{\l,\mu,\nu}$. As a result, up to a normalization factor, the topological vertex counts the configurations of an infinite cube with boxes removed at the corner, effectively describing a melting crystal. This analogy with the melting crystal follows from the rewriting of the topological vertex as a correlator of operators acting in the free boson Fock space $\CF_0$. In order to point out the connection with the intertwiner $\Phi(v)$, we briefly sketch their derivation. It starts from the well-known formula of the topological vertex written in terms of skew-Schur polynomials
\begin{equation}\label{def_top_ver}
C_{\l,\mu,\nu}=q^{-\k(\l)/2-\k(\nu)/2}s_{\nu'}(q^{-\rho})\sum_\eta s_{\l'/\eta}(q^{-\rho-\nu})s_{\mu/\eta}(q^{-\rho-\nu'}).
\end{equation} 
The argument $q^{-\rho-\nu}$ of the skew-Schur functions indicates the evaluation of the polynomial at $(x_1,x_2,\cdots)=(q^{-\nu_1+1/2},q^{-\nu_2+3/2},\cdots)$. The rewriting of the topological vertex is based on the realization of skew-Schur polynomials as the matrix elements of the operators $\G_\pm(x)$ in the Schur basis,
\begin{equation}\label{slm_Gamma}
s_{\l'/\eta}(x)=\bra{\l'}\G_-(x)\ket{\eta},\quad s_{\mu/\eta}(x)=\bra{\eta}\G_+(x)\ket{\mu},\quad \Gamma_\pm(x) =\exp\left(\sum_{k>0} \frac{1}{k}p_k(x)\ J_{\pm k} \right).
\end{equation}
We refer the reader to the appendix of ref. \cite{Bourgine2021} for a short derivation of these well-known formulas. Then, the sum over partitions $\eta$ in the expression \ref{def_top_ver} of the topological vertex can be performed using the closure relation of the Schur basis, leading to
\begin{equation}
C_{\l,\mu,\nu}=q^{-\k(\l)/2-\k(\nu)/2}s_{\nu'}(q^{-\rho})\langle \lambda'|\Gamma_-(q^{-\nu'-\rho})\Gamma_+(q^{-\nu-\rho})|\mu\rangle.
\end{equation}
In order to build the 3d partitions of the melting crystal, the two operators $\G_\pm(x)$ need to be exchanged inside the correlator, it produces an extra factor that is easily computed from their bosonic expression,
\begin{equation}\label{C_Gamma}
C_{\l,\mu,\nu}=(-1)^{|\nu|}q^{-\k(\l)/2}Z(q)^{-1}(s_{\nu'}(q^{-\rho}))^{-1}\langle \lambda'|\Gamma_+(q^{-\nu'-\rho})\Gamma_-(q^{-\nu-\rho})|\mu\rangle,
\end{equation}
where $Z(q)$ is MacMahon's generating function of plane partitions,
\begin{equation}
Z(q)=\sum_{\pi\in\text{P.P.}}q^{|\pi|}=\prod_{n=1}^\infty(1-q^n)^{-n}.
\end{equation}
From its bosonic expression \ref{def_Phi}, the intertwiner $\Phi(v)$ at $v=q^{-1/2}$ is identified with the operator inside the correlators after normal-ordering,
\begin{equation}\label{AFS_ORV}
\Phi_\nu(q^{-1/2})=t_\nu\G_-(q^{-\nu-\rho})\G_+(q^{-\nu'-\rho})\implies C_{\l,\mu,\nu}=q^{-\k(\l)/2-\k(\nu)/2}\langle \lambda'|\Phi_\nu(q^{-1/2})|\mu\rangle.
\end{equation}
Once we set $v=q^{-1/2}$, taking into account the various constraints, only two weights are free to choose in the intertwining relation \ref{AFS_lemma}, e.g. $u_{11}$ and $v_{11}$, and the others are fully determined: $u_{10}=-u_{11}/v_{11}$, $v_{10}=v_{11}$, $u_{01}=q^{1/2}v_{11}$, $v_{01}=q^{1/2}v_{11}/u_{11}$. In the following, we denote for simplicity $\Phi_\nu=\Phi_\nu(q^{-1/2})$. In fact, other values of the parameter $v$ can be considered by insertion of the operator $Q^{L_0}$ that acts on the modes as $Q^{L_0}J_kQ^{-L_0}=Q^{-k}J_k$ and so $Q^{L_0}\Phi_\l(v)Q^{-L_0}=\Phi_\l(Qv)$. Coming back to the vacuum component, we have found that $\Phi_\vac=\G_-(q^{-\rho})\G_+(q^{-\rho})$ and thus the exchange relation \ref{AFS_proj} does indeed coincide with the shift symmetries derived by Nakatsu and Takasaki.

\subsection{Derivation of the tau function}
The starting point for the construction of the KP tau function is MacMahon's generating function that can be written as a sum of Schur polynomials using the Cauchy identity. This expression is then deformed by the introduction of the time parameters $\bst=(t_k)$ coupled to the eigenvalues \ref{act_W0k} of the operator $W_{0,-k}$,\footnote{For convenience, the overall factor $Q^{s(s+1)}$ has been removed with respect to the formula (4.4) given in \cite{Takasaki2018}.}
\begin{equation}\label{def_Zq}
Z(q)=\sum_\l (s_\l(q^{-\rho}))^2\quad\to\quad Z(q,s,\bst)=\sum_\l(s_\l(q^{-\rho}))^2 Q^{|\l|}e^{\sum_{k>0}t_k\phi_k(\l,s)}.
\end{equation}
We do not indicate the $Q$-dependence as we can think of it as an extra time parameter $t_0=\log Q$. To show that the quantity on the right is a tau function of the KP hierarchy, we need to rewrite it as a bosonic correlator in the canonical form \ref{canonical}.

The first step is similar to the rewriting of the topological vertex in the melting crystal picture, namely the Schur functions are replaced by bosonic correlators using the formulas \ref{slm_Gamma}, and the time dependence follows from the diagonal action of  $W_{0,-k}$ and $L_0$ on the Schur states,
\begin{equation}
Z(q,s,\bst)=\sum_\l\bra{\vac}\G_+(q^{-\rho})\rho_{q^{-s},v}^{(D)}(e^{H(\bst)})\ket{\l}\bra{\l}Q^{L_0}\G_-(q^{-\rho})\ket{\vac},\quad H(\bst)=\sum_{k>0}t_kW_{0,-k}.
\end{equation}
From our previous remark, it is clear that the function $Z(q,s,\bst)$ does not depend on the weight $v$ and we take $v=1$. The sum over partitions $\l$ is eliminated using the closure relation of the Schur basis in order to write
\begin{equation}
Z(q,s,\bst)=\bra{\vac}\G_+(q^{-\rho})\rho_{q^{-s},1}^{(D)}(e^{H(\bst)})Q^{L_0}\G_-(q^{-\rho})\ket{\vac}.
\end{equation}

In the second step, the extra vertex operators $\G_\pm(q^{-\rho})$ are introduced on the left/right of the correlators. Since these operators are built purely upon either positive or negative modes, they give no extra contribution. Based on our previous discussion,  we now understand that the reason behind this clever trick is to reconstruct the vacuum component of the intertwiner,
\begin{equation}
Z(q,s,\bst)=\bra{\vac}\Phi_{\vac}\rho_{q^{-s},v}^{(D)}(e^{H(\bst)})Q^{L_0}\Phi_{\vac}\ket{\vac}.
\end{equation}
It allows us to use the exchange relation \ref{AFS_proj} for the exponential of the evolution operator $H(\bst)$.
\begin{equation}\label{rel_xchg}
\rho^{(D)}_{-u,1}(\CT\cdot e^{H(\bst)})\Phi_{\vac}=\Phi_{\vac}\rho^{(D)}_{u,1}(e^{H(\bst)}).
\end{equation} 
Computing the action of the automorphism $\CT$ on these operators with the help of the equation \ref{transfo_ST}, we can rewrite the amplitude as
\begin{equation}
Z(q,s,\bst)=e^{\sum_{k>0}t_k\frac{q^k}{1-q^k}}\bra{\vac}\rho_{-q^{-s},1}^{(D)}\left(e^{\sum_{k>0}t_kq^{-k^2/2}W_{k,-k}}\right)\Phi_{\vac}Q^{L_0}\Phi_{\vac}\ket{\vac}.
\end{equation} 

The last step consists in transforming the generators $W_{k,-k}$ into the Heisenberg modes $J_k=\bW_{k,0}$. It is done using the framing operators since the specialization of the formula \ref{tW0_action} at $\a=1$ and $(m,n)=(k,-k)$ reads $q^{\tW_0}\rho_{u,1}^{(D)}(W_{k,-k})q^{-\tW_0}=u^{-k}q^{k^2/2}J_k$. The framing operator can be introduced on the left/right of the correlator for free since its action on the vacuum state is trivial. In this way, we find the desired result
\begin{equation}\label{tau_sd}
Z(q,s,\bst)=e^{\sum_{k>0}t_k\frac{q^k}{1-q^k}}\bra{\vac}e^{\sum_{k>0}(-q^s)^kt_kJ_k}g\ket{\vac},\quad g=q^{\tW_0}\Phi_{\vac}Q^{L_0}\Phi_{\vac}q^{\tW_0}.
\end{equation}
Then, the fact that $Z(q,s,\bst)$ is a tau function of the KP hierarchy follows from the fact that $g\otimes g$ commutes with $\Psi=\sum_r\psi_r\otimes\bpsi_{-r}$. This type of operators were called the group-like elements of $GL(\infty)$ in \cite{Alexandrov2012}. This property is sufficient to ensure that $Z(q,s,\bst)$ obeys the Hirota equation. Let us stress that the bosonic formula for $\Phi_\vac$ and the fermionic realization of the framing operator through \ref{def_JLW} are essential to show the group-like property. The fact that the hierarchy is of KP-type follows from the possibility to move the time dependency to the right of the correlators. We refer to the original papers \cite{Nakatsu2007,Takasaki2013,Takasaki2018} for a more thorough discussion that also includes the Lax formalism.

We conclude this section with a short remark. One may wonder what would happen if we considered the other vertical components $\Phi_\nu$ of the intertwining operator instead of the vacuum component. In this case, one can show that the exchange relation \ref{AFS_proj} is satisfied only for the generators $W_{m,n}$ with the index $n<|\nu|$. As a consequence, we need to restrict ourselves to the times $t_k$ with $k>|\nu|$. Replacing $\Phi_\vac$ with $\Phi_\nu$ in the operator $g$ and turning off the lower times, we find another tau function that reads
\begin{equation}
Z_{\nu}(q,s,\bst)=t_\nu^2\sum_\l s_\l(q^{-\rho-\nu})s_\l(q^{-\rho-\nu'}) Q^{|\l|}e^{\sum_{k>|\nu|}t_k\phi_k(\l,s)}.
\end{equation} 
This tau function appears to be a particular case of the generating functions considered in \cite{Nakatsu2018}. It is naturally associated to the topological vertex $C_{\vac,\vac,\nu}$ that enumerates plane partitions with fixed asymptotics $\nu$ in one direction \cite{ORV}.

\section{Refinement and quantum toroidal algebra}
In this section, we attempt to generalize the construction of a tau function to the refined melting crystal introduced in the context of topological strings in \cite{Iqbal2007}. The main idea is to consider a counting function of plane partitions where boxes have a different weight $q_{\sAcube}=q,t$ depending on their location,\footnote{We refer to the appendix A of ref. \cite{Iqbal2007} for the exact prescription.}
\begin{equation}\label{def_Ztq}
Z(t,q)=\sum_{\pi}\prod_{\sAcube\in\pi}q_{\sAcube}=\prod_{i,j=1}^\infty(1-t^{i}q^{j-1})^{-1}.
\end{equation} 
Using the Cauchy identity, the double product can be written as a sum over Macdonald polynomials,
\begin{equation}
Z(t,q)=\sum_\l \dfrac{(P_\l(t^{-\rho}))^2}{\la P_\l,P_\l\ra_{q,t}},
\end{equation}
it reduces to the expression \ref{def_Zq} of $Z(q)$ in the limit $t=q$ since Macdonald polynomials reduce to Schur polynomials and their norm \ref{norm_Mac} tend to one. However, in order to be able to employ the intertwiner $\Phi(v)$, instead of its dual $\Phi^\ast(v)$, it is more convenient to generalize the formulas \ref{def_Zq} as
\begin{equation}\label{def_Z_Q_t}
Z(t,q)=\sum_\l \dfrac{(\iota P_\l(t^{-\rho}))^2}{\la P_\l,P_\l\ra_{q,t}}\quad\to\quad Z(Q,\bst,t,q)=\sum_\l Q^{|\l|}\dfrac{(\iota P_\l(t^{-\rho}))^2}{\la P_\l,P_\l\ra_{q,t}}e^{\sum_{k>0}t_k \Phi_k(\l,u)},
\end{equation}
where we used the involution $\iota:p_k(x)\to-p_k(x)$ acting on the ring of symmetric polynomials by reversing the sign of elementary power sums.\footnote{Both $P_\l(x)$ and $\iota P_\l(x)$ have the same Cauchy identity, which provides two different ways of writing the double infinite product \ref{def_Ztq}. Note also that in the limit $t=q$, $\iota P_\l(x)\to(-1)^{|\l|}s_{\l'}(x)$ and the first equality in \ref{def_Z_Q_t} reproduces again the expression \ref{def_Zq} of $Z(q)$ upon the replacement $\l\to\l'$ in the summation.} In order to define properly the deformed quantities $\Phi_k(\l,u)$, we need to generalize the algebraic description to the refined case. The relevant algebra is the quantum toroidal algebra of $\mathfrak{gl}(1)$, it depends on two parameters $(q_1,q_2)$ that are identified with the parameters $(q,t^{-1})$ of the Macdonald polynomials. It reduces to the quantum $\Winf$ algebra in the limit $t\to q$ which corresponds to the self-dual limit of the omega-background $\e_1+\e_2\to0$. Moreover, it also has a representation acting on the module $\CF_0$, called the \textit{Fock (or horizontal) representation} \cite{Feigin2009a,feigin2011quantum}, which we will review shortly below.

To obtain the bosonic expression \ref{canonical} for the amplitude $Z(Q,\bst,t,q)$, two main ingredients are needed: an exchange relation generalizing \ref{rel_xchg} and a \textit{framing operator} that will play the role of $q^{\tW_0}$. Like before, the exchange relation will be obtained as a projection of the AFS intertwining relation \cite{AFS} on the vacuum component of the intertwiner in the vertical direction. On the other hand, the framing operator will be constructed by considering the refined framing factors \cite{Taki2007,Iqbal2007}. We will then show that this operator obeys an equivalent of the algebraic property \ref{tW0_action}.

\subsection{Quantum toroidal $\mathfrak{gl}(1)$ algebra}
\subsubsection{Definition}
We review briefly here the definition of the quantum toroidal $\mathfrak{gl}(1)$ algebra. We mostly follow the notations and conventions of \cite{Bourgine2018a}, and refer to \cite{Awata2016,Awata2016a,Mironov2016,Bourgine2017b} for more details on the correspondence with the $(p,q)$-brane construction of topological strings amplitudes.

The algebra is usually formulated in terms of four Drinfeld currents,
\begin{equation}\label{DIM_currents}
x^\pm(z)=\sum_{k\in\mathbb{Z}}z^{-k}x^\pm_k,\quad \psi^\pm(z)=\sum_{k\geq0}z^{\mp k}\psi_{\pm k}^\pm.
\end{equation} 
They satisfy a set of exchange relations that can be found, e.g. in \cite{Bourgine2017b,Bourgine2018a}, but we prefer to work here directly with the modes $x_{k}^\pm$, $\psi_{\pm k}^\pm$. The subalgebra generated by the elements $\psi_{\pm k}^\pm$ is the analogue of the Cartan subalgebra of quantum affine algebras, it has an alternative formulation in terms of modes $a_k$ defined by exponentiation,
\begin{equation}\label{def_ak}
\psi^\pm(z)=\psi_0^\pm\exp\left(\pm\sum_{k>0}z^{\mp k}a_{\pm k}\right),
\end{equation}
and satisfying a twisted Heisenberg algebra. The algebra has only two parameters $(q_1,q_2)$, but it is useful to introduce a third one through the relation $q_1q_2q_3=1$. We also introduce the shortcut notation $\g=q_3^{1/2}=(q_1q_2)^{-1/2}$. The algebra has two central charges $(c,\bc)$, the second one entering through the zero modes of the Cartan currents $\psi_0^\pm=\g^{\mp\bar c}$. The modes of the currents satisfy the commutation relations
\begin{align}
\begin{split}\label{com_ak}
[a_k,a_l]=(\g^{kc}-\g^{-kc})c_k\d_{k+l},\quad [a_k,x_l^\pm]=\pm\g^{\mp |k|c/2}c_k x_{l+k}^\pm,\\
[x^+_k,x^-_l]=\left\{
\begin{array}{l}
\k\g^{(k-l)c/2}\psi^+_{k+l},\quad k+l>0\\
\k\g^{(k-l)c/2}\psi_0^+-\k\g^{-(k-l)c/2}\psi_0^-,\quad k+l=0\\
-\k\g^{-(k-l)c/2}\psi^-_{k+l},\quad k+l<0,\\
\end{array}
\right.
\end{split}
\end{align}
with the coefficients
\begin{equation}\label{def_ck}
\k=\dfrac{(1-q_1)(1-q_2)}{(1-q_1q_2)},\quad c_k=-\dfrac1{k}\prod_{\a=1,2,3}(1-q_\a^k).
\end{equation}
They form a Hopf algebra with the Drinfeld coproduct
\begin{align}
\begin{split}\label{coproduct}
&\D(x_k^+)=x^+_k\otimes 1+\sum_{l\geq0}\g^{-(c\otimes1)(k+l/2)}\ \psi^-_{-l}\otimes x^+_{k+l},\quad \D(x_k^-)=\sum_{l\geq0}\g^{-(1\otimes c)(k-l/2)}\ x^-_{k-l}\otimes \psi^+_l+1\otimes x_k^-,\\
&\D(a_k)=a_k\otimes \g^{-|k|c/2}+\g^{|k|c/2}\otimes a_k,\quad \D(c)=c\otimes 1+1\otimes c,\quad \D(\bc)=\bc\otimes1+1\otimes \bc.
\end{split}
\end{align}

\paragraph{Automorphisms} The algebra is known to possess the group of automorphisms $\text{SL}(2,\mZ)$ generated by the elements $\CS$ and $\CT$. The action of the automorphism $\CT$ on the modes of the Drinfeld currents can be expressed easily: it leaves the Cartan modes $a_k$ invariant and acts as
\begin{equation}
\CT\cdot x_k^\pm=x_{k\mp1}^\pm,\quad \CT\cdot(c,\bc)\to (c,\bc+c).
\end{equation} 
The automorphism $\CS$ has been uncovered by Miki in \cite{Miki2007}. It is of order four, and is defined uniquely by its action on the modes $x_0^\pm$, $a_{\pm1}$, namely 
\begin{equation}\label{Miki_init}
a_1\to(\g-\g^{-1})x_0^+\to -a_{-1}\to -(\g-\g^{-1})x_0^-\to a_1,
\end{equation}
and the central elements $(c,\bc)\to(-\bc,c)$. The explicit transformation formulas of the modes $x_k^\pm$ and $\psi_{\pm k}^\pm$ are useful here, but, since they are more complicated, we decided to confine them to the appendix \ref{AppA} to avoid introducing too many notations. Let us only define the notation $b_k=\CS\cdot a_k$ for the $\CS$-dual Cartan modes.

\paragraph{Self-dual limit} In the self-dual limit $(q_1,q_2)\to(q,q^{-1})$, the modes of the quantum toroidal $\gl(1)$ algebra satisfy the commutation relations of the quantum $\Winf$ algebra. The identification goes as follows,
\begin{align}
\begin{split}
&\dfrac{x_k^+}{1-q_1}\to q^{k/2}W_{k,1}+\d_{k,0}\dfrac{c_2}{1-q},\quad \dfrac{x_k^-}{1-q_2}\to q^{-k/2}W_{k,-1}+\d_{k,0}\dfrac{c_1}{1-q^{-1}},\\
&\dfrac{ka_k}{(q_1^{k/2}-q_1^{-k/2})(q_3^{k/2}-q_3^{-k/2})}\to W_{k,0}-\dfrac{c_2}{1-q^k},\quad (c,\bc)\to(c_1,-c_2).
\end{split}
\end{align}
Thus, the roles of $W_{k,0}$, $W_{0,k}$ and $W_{k,-k}$ is played in the refined case by $a_k$, $b_{-k}$ and $\CT\cdot b_{-k}$ respectively.\footnote{These modes correspond to the three patches in \cite{Aganagic2003}, as can be seen from their fermionic realization
\begin{equation}
\bW_{k,0}=\oint{\dfrac{dz}{2i\pi}z^k:\bpsi(z)\psi(z):},\quad\bW_{0,k}=\oint{\dfrac{dz}{2i\pi}:\bpsi(z)\psi(e^{k g_\text{str}}z):},\quad\bW_{k,-k}=\oint{\dfrac{dz}{2i\pi}z^k:\bpsi(z)\psi(e^{-k g_\text{str}}z):}.
\end{equation}}

\subsubsection{Horizontal representation}
The role previously devoted to the Dirac representation $\rho_{u,v}^{(D)}$ will now be played by the horizontal representation \cite{Feigin2009a}. This representation has also the levels $(1,0)$ and acts on the free boson Fock space $\CF_0$. It has a weight $u\in \mC^\times$ and will be denoted $\rho_u^{(1,0)}$. In this representation, the Drinfeld currents take the form of vertex operators defined upon the Heisenberg modes $J_k$ representing the Cartan modes $a_k$,
\begin{align}
\begin{split}\label{q-osc}
&\rho^{(1,0)}_u(a_k)=\dfrac{\g^{k/2}}{k}(\g^k-\g^{-k})(1-q_2^k)J_k,\quad \rho^{(1,0)}_u(a_{-k})=\dfrac{\g^{k/2}}{k}(\g^k-\g^{-k})(1-q_1^k)J_{-k},\quad (k>0),\\
&\rho^{(1,0)}_u(x^\pm(z))=u^{\pm1}\exp\left(\pm\sum_{k>0}\dfrac{z^{k}}{k}\g^{(1\mp1)k/2}(1-q_1^k)J_{-k}\right)\exp\left(\mp\sum_{k>0}\dfrac{z^{-k}}{k}\g^{(1\mp1) k/2}(1-q_2^k)J_k\right).
\end{split}
\end{align}
Using the isomorphism between $\CF_0$ and the ring of symmetric polynomials, the action of the modes $a_k$ on the Macdonald states $\ket{P_\l}$ corresponds to either a multiplication by the power sums $p_k$ or the derivation $\p/\p p_{k}$ depending on the sign of $k$. Since the power sum $p_1$ coincides with the elementary symmetric polynomial $e_1$, the action of $a_{\pm1}$ is deduced from the Pieiri rules obeyed by Macdonald polynomials,
\begin{equation}\label{Pieri}
\rho^{(1,0)}_u(a_1)\ket{P_\l}=\sum_{\sAbox\in R(\l)}r_\l^{(-)}(\Abox)\ket{P_{\l-\sAbox}},\quad \rho^{(1,0)}_u(a_{-1})\ket{P_\l}=\sum_{\sAbox\in A(\l)}r_\l^{(+)}(\Abox)\ket{P_{\l+\sAbox}}.
\end{equation} 
We denoted here $A(\l)$ and $R(\l)$ the sets boxes that can be added to/removed from the Young diagram of $\l$. The coefficients $r_\l^{(\pm)}(\Abox)$ depend on the choice of normalization for the modes and states, they can be computed explicitly but we will not need their expression here.

Finally, using Miki's automorphism, we can also determine the action of the modes $b_k$ on the Macdonald states \cite{Bourgine2018a}. It is used here to define the quantities $\Phi_k(\l,u)$ coupled to the times $t_k$,
\begin{align}
\begin{split}\label{action_bk}
&\rho^{(1,0)}_{u}(b_{-k})\ket{P_\l}=\Phi_{k}(\l,u)\ket{P_\l},\\
&\Phi_{k}(\l,u)=(-1)^k\g^{k}u^{-k}c_k\left(\sum_{(i,j)\in\l}q_1^{-(i-1)k}q_2^{-(j-1)k}-\dfrac1{(1-q_1^{-k})(1-q_2^{-k})}\right)\ket{P_\l}.
\end{split} 
\end{align}
Since this action is diagonal, it does not depend on the choice of normalization for the Macdonald states. These operators are proportional to the Macdonald operators known to act diagonally on the polynomials $P_\l$. In the self-dual limit, 
\begin{equation}
\dfrac{k\Phi_k(\l,q^{-s})}{(q_1^{k/2}-q_1^{-k/2})(q_3^{k/2}-q_3^{-k/2})}\to -(-1)^{k}q^{-k/2}\left[\phi_k(\l,s)+\dfrac1{1-q^{-k}}\right],
\end{equation} 
and the deformed amplitude $Z(Q,\bst,t,q)$ reduces to the tau function \ref{tau_sd} (up to a rescaling of the times parameters). Note, however, that the second term in the bracket is responsible for an extra exponential factor in the formula \ref{tau_sd}, it will not be present in the refined case.

\subsubsection{Intertwiner and exchange relation}
The intertwiner constructed by Awata, Feigin and Shiraishi in \cite{AFS} intertwines the representation of levels $(1,n+1)$ and the tensor product of two representations with levels $(0,1)$ and $(1,n)$. Here, we only need to consider the case $n=0$. Moreover, the representations of levels $(1,1)$ and $(0,1)$ can be obtained from the horizontal representation $(1,0)$ using the automorphisms $\CS$ and $\CT$ described above. From the analysis of the transformation of representations performed in \cite{Bourgine2018a}, the AFS intertwining equation can be rewritten in the form
\begin{equation}\label{AFS_lemma_ref}
\rho_{uv}^{(1,0)}(\CT\cdot e)\Phi(v)=\Phi(v)\left(\rho_v^{(1,0)}\circ\CS\otimes\rho_u^{(1,0)}\ \D(e)\right),
\end{equation}
for any element $e$ of the quantum toroidal $\gl(1)$ algebra. The solution of this equation has been found in \cite{AFS}, it can be expanded over the vertical components as a sum of vertex operators,
\begin{align}
\begin{split}
&\Phi(v)=\sum_\l\Phi_\l(v)\dbra{P_\l},\quad \Phi_\l(v)=t_\l:\Phi_\vac(v)\prod_{(i,j)\in\l}\eta(vq_1^{i-1}q_2^{j-1}):,\\
&\text{with}\quad \Phi_\vac(v)=\exp\left(-\sum_{k>0}\dfrac{v^k}{k(1-q_2^k)}J_{-k}\right)\exp\left(\sum_{k>0}\dfrac{q_3^{-k}v^{-k}}{k(1-q_1^k)}J_k\right).
\end{split}
\end{align}
The operator $\eta(z)=\rho_1^{(1,0)}(x^+(z))$ coincides with the representation of the current $x^+(z)$ given in \ref{q-osc}. The normalization factor $t_\l$ is not important here as we only consider the vacuum component, and we can always set $t_\vac=1$. In the melting crystal formalism, the vacuum component corresponds to
\begin{equation}
\Phi_{\vac}(v)=\G_-(vt^{1/2-\rho})\G_+(v^{-1}\g^{-2}q^{-1/2-\rho}).
\end{equation} 

To derive the exchange relation, we exploit the fact that $\CS\cdot x_k^-$ annihilates the vacuum state $\dket{\vac}$ in the vertical channel. This fact follows from the application of Miki's automorphism to map the vertical representation of levels $(0,1)$ to the horizontal one, since the vertical action of $x_k^-$ annihilates the vacuum \cite{Bourgine2018a}. However, to derive an exchange relation for the modes $b_{-k}$, there is small difficulty coming from the fact that their coproduct involves an infinite sum. For instance, for $b_{-1}=(\g-\g^{-1})x_0^-$, we have
\begin{equation}
(\g-\g^{-1})^{-1}\D(b_{-1})=\sum_{k\geq 0}\g^{k(1\otimes c)/2}x_{-k}^-\otimes\psi_k^++1\otimes x_0^-.
\end{equation}
Here, the AFS equation \ref{AFS_lemma_ref} does simplify into an exchange relation because the vertical action of all the terms $x_{-k}^-\otimes\psi_k^+$ vanishes. After the projection of the resulting equation on the vertical vacuum component, we find
\begin{equation}
\rho_{uv}^{(1,0)}(\CT\cdot b_{-k})\Phi_\vac(v)=\Phi_{\vac}(v)\rho_u^{(1,0)}(b_{-k})
\end{equation} 
for $k=1$. The proof for higher $k>0$ follows from the same properties. Since it is a little technical, we kept it in appendix \ref{AppA}. This exchange relation implies for $H(\bst)=\sum_{k>0}t_kb_{-k}$,
\begin{align}
\begin{split}\label{rel_xchg_ref}
&\rho_{uv}^{(1,0)}(\CT\cdot e^{H(\bst)})\Phi_\vac(v)=\Phi_{\vac}(v)\rho_u^{(1,0)}(e^{H(\bst)}),\\
&\rho_{uv}^{(1,0)}(e^{H(\bst)})\Phi_\vac(v)=\Phi_{\vac}(v)\rho_u^{(1,0)}(\CT^{-1}\cdot e^{H(\bst)}).
\end{split}
\end{align}
The second exchange relation is also derived in appendix \ref{AppA}, it is obtained from the AFS relation \ref{AFS_lemma_ref} applied to $e=\CT^{-1}\cdot b_{-k}$

\subsection{Deforming the tau function}
To avoid cluttering the formulas, we omit the dependence in the variables $(q,t)$ in this section. In order to construct the bosonic expression for the amplitude $Z(Q,\bst)$ defined in \ref{def_Z_Q_t}, we need the expression of the matrix elements of the intertwiner $\Phi(v)$ in the Macdonald basis,\footnote{Since $p_k(t^{\pm\rho})=\pm1/(t^{k/2}-t^{-k/2})$ with $|t|^{\pm1}>1$ for convergence, we need to replace $P_\l(t^\rho)\to \iota P_\l(t^{-\rho})$ in the expression (4.13) of ref. \cite{AFS} when $|t|<1$.}
\begin{equation}
\bra{\vac}\Phi_\vac(q^{1/2}v)\ket{P_\l}=v^{-|\l|}\g^{-|\l|}\iota P_\l(t^{-\rho}),\quad \bra{P_\l}\Phi_\vac(q^{1/2}v)\ket{\vac}=v^{|\l|}\g^{-|\l|}\iota P_\l(t^{-\rho}).
\end{equation}
These formulas are used to replace the Macdonald polynomials in \ref{def_Z_Q_t} with bosonic correlators. Then, the time dependence is produced using the diagonal action \ref{action_bk} of the modes $b_{-k}$ on the Macdonald basis. Finally, the summation of the Young diagrams $\l$ is performed using the closure relation of the Macdonald basis, and we find
\begin{equation}
Z(\g^{-2}Q,\bst)=\bra{\vac}\Phi_\vac(v)\rho_u^{(1,0)}(e^{H(\bst)})\Phi_\vac(Qv)\ket{\vac}.
\end{equation} 
Note that this quantity is actually independent of $v$, and the dependence in $u$ can be eliminated by a rescaling of the times $t_k\to u^k t_k$. The second intertwiner can be replaced by $\Phi_\vac(Qv)=Q^{L_0}\Phi_\vac(v)Q^{-L_0}$ where $L_0$ is the Fock space operator introduced in \ref{def_JLW}. Since Macdonald polynomials, just like Schur polynomials, are homogeneous polynomials of degree $|\l|$, it acts diagonally on the Macdonald basis, $L_0\ket{P_\l}=|\l|\ket{P_\l}$.\footnote{We can also see $L_0$ as the representation of the grading operator $d$ for the quantum toroidal algebra (see \cite{Bourgine2018a})
\begin{equation}
\rho_u^{(1,0)}(d)=\sum_{k>0}J_{-k}J_k=\sum_{r\in\mZ+1/2}r:\bpsi_{-r}\psi_r:=L_0.
\end{equation}}

We proceed to move the time dependence to the left. The first step is performed using the exchange relation \ref{rel_xchg_ref}, it gives
\begin{equation}
Z(\g^{-2}Q,\bst)=\bra{\vac}\rho_{uv}^{(1,0)}(\CT\cdot e^{H(\bst)})\Phi_\vac(v)\Phi_\vac(Qv)\ket{\vac}.
\end{equation} 
For the next step, we need to define the framing operator. In the refined case, the framing factor is modified into $f_\l=q^{n(\l')}t^{-n(\l)}$ with $n(\l)=\sum_{(i,j)\in\l}(i-1)$ and $n(\l')=\sum_{(i,j)\in\l}(j-1)$. It prompts us to define the framing operator $F$ as a diagonal operator on the Macdonald states, with eigenvalues
\begin{equation}\label{def_F}
F\ket{P_\l}=F_\l\ket{P_\l},\quad F_\l=\prod_{(i,j)\in\l}q_1^{i-1}q_2^{j-1}=q_1^{n(\l)}q_2^{n(\l')}.
\end{equation}
In the limit $t=q$, this operator tends to $q^{-W_0/2}$, the shift $L_0$ is missing because we treat the $v$-dependence differently here. Like the operator $q^{\tW_0}$ in the self-dual case, the operator $F$ is deeply connected to the automorphism $\CT$, i.e.
\begin{align}
\begin{split}\label{id_T_bk}
&\rho_{u}^{(1,0)}(\CT\cdot b_{-k})=(-1)^{k+1}u^{-k}\g^{k/2}\ F\rho_{u}^{(1,0)}(a_k)F^{-1},\quad k\in\mZ,\\
&\rho_{u}^{(1,0)}(\CT^{-1}\cdot b_{-k})=(-1)^{k}u^{-k}\g^{k/2}\ F^{-1}\rho_{u}^{(1,0)}(a_{-k})F,\quad k\in\mZ.
\end{split}
\end{align}
The derivation of these two identities is a bit involved as it makes use of Miki's automorphism, it is presented in appendix \ref{AppA}. Using the first identity, and the horizontal representation \ref{q-osc} of the modes $a_k$, we can write the partition function in the form
\begin{equation}\label{Ztq_corr}
Z(\g^{-2}Q,\bst)=\bra{\vac}e^{\sum_k\frac1k\t_kJ_k}g\ket{\vac},\quad\text{with}\quad g=F^{-1}\Phi_\vac(v)\Phi_\vac(Qv)F^{-1},
\end{equation} 
and the rescaled times $\t_k=(-uv)^{-k}(1-q_2^k)(1-q_3^k)t_k$. This equation is the main result of this section. Unfortunately, we were not able to show that the operator $g$ is a group like element of $GL(\infty)$. Group like elements form a monoid, and we can examine the decomposition of $g$ into its elementary factors. The vacuum component of the intertwiner is still a group like element, despite the asymmetry between positive and negative modes as any operator of the form $:e^{\sum_{k\in\mZ^\times}t_kJ_k}:$ is group-like. On the other hand, we could not show that $F$ is group-like, and strongly suspect that it is not from a perturbative analysis of the Hirota equation. As we shall explain in the next section, it is likely that the operator $\Psi$ has to be replaced by a different operator.

We conclude with another remark. Using the second exchange relation in \ref{rel_xchg_ref} together with the second identity in \ref{id_T_bk}, the exponential of $H(\bst)$ can be moved to the right in the correlator instead,
\begin{equation}
Z(\g^{-2}Q,\bst)=\bra{\vac}ge^{-\sum_{k>0}\bar\t_kJ_{-k}}\ket{\vac},
\end{equation} 
with $\bar\tau_k=(-u/Qv)^{-k}(1-q_1^k)(1-q_3^k)t_k$. Unlike in the case of the KP hierarchy, the times variables are not equal but they obey a simple scaling relation,
\begin{equation}\label{rel_tau}
\bar\t_k=\dfrac{1-q_1^k}{1-q_2^k}Q^kv^{2k}\t_k.
\end{equation} 


\section{Discussion}
Our main result is the observation of a relation between the tau function of an integrable hierarchy, the intertwining operator of a quantum algebra, and the framing operator. The fundamental role of the $\text{SL}(2,\mZ)$ group of automorphisms in this description has been emphasized as it is deeply related to both intertwiner and framing operator. Our observation offers the possibility to extend the correspondence between topological strings theory and integrable hierarchies in several new directions. The most obvious one is to consider more involved toric diagrams by exploiting the gluing rules of the topological vertex \cite{Mironov2016,Awata2016a,Bourgine2017b}. The next simplest toric diagram describes the resolved conifold and the corresponding time-deformed amplitude was shown to be a tau function of a different reduction of the Toda hierarchy called the Ablowitz-Ladik hierarchy \cite{Brini2010,Takasaki2013}. This result could be reproduced within our algebraic formalism provided that we introduce the dual intertwiner $\Phi^\ast$, also constructed in \cite{AFS}, that is expected to enjoy a similar exchange relation.

The trinion theories $T_N$ provide another set of interesting toric diagrams \cite{Bao2013,Hayashi2013,Coman2019}.\footnote{The author would like to thank E. Pomoni for drawing his attention to this family of theories.} The algebraic object obtained by gluing intertwiners according to these diagrams is also an intertwiner but it involves representations of the quantum toroidal algebra with higher levels, namely $(N,N)$, $(N,0)$ and $(0,N)$. The intertwining relation projected on the vertical component is expected to produce an exchange relation similar to \ref{rel_xchg_ref}. Furthermore, the framing operator can be easily generalized to representations $(0,N)$ by taking the tensor product of the $(0,1)$ framing operator defined in this paper. Thus, our construction should apply in this case as well, and the corresponding time-deformed amplitude should be the tau function of an integrable hierarchy in the self-dual limit.

The second part of the paper is an attempt to define a refined tau function from the natural deformation of the algebraic objects, lifting them from the quantum $\Winf$ algebra to the quantum toroidal $\gl(1)$ algebra. We ended up with the bosonic expression \ref{Ztq_corr} for the refined amplitude, but we were unable to show that the refined framing operator entering in the operator $g$ is a group-like element for $GL(\infty)$. Since the Dirac fermion plays no role in the representation theory of the quantum toroidal algebra, we suspect that the Casimir operator $\Psi=\sum_r\psi_r\otimes\bpsi_{-r}$ is replaced by a different operator, just like in the Kac-Wakimoto construction. At the moment, it is not clear what this operator should be. In this respect, the Kac-Wakimoto construction for the toroidal algebra realized in \cite{Ikeda2000} might be a good source of inspiration.

As an intermediate step in the deformation of the fermionic structure, one could focus on the $q=0$ limit in which Macdonald polynomials reduce to Hall-Littlewood polynomials. In \cite{Jing1991}, Jing introduced certain vertex operators as a $t$-deformation of the Dirac fermion. These operators might be used to deform the Hirota equation, or the Lax formalism.

The algebraic description of topological string theory has been extended to different algebras and geometric backgrounds. Some of these algebras should possess an $\text{SL}(2,\mZ)$ subgroup of automorphisms, like the quantum toroidal $\gl(p)$ algebras \cite{Awata2017}, their elliptic deformations \cite{Foda2018,Ghoneim2020} or even the fully deformed algebra of \cite{Bourgine:2019phm}. In all these cases, we expect our construction to apply, producing tau functions of different integrable hierarchies in specific limits. 

Going in the other direction, one might try to build a vertex operator from known integrable hierarchies. For instance, the quantum algebra associated to BKP, CKP and DKP hierarchies \cite{Date1981,Jimbo1983} are known to be orbifolds of the quantum $\Winf$ algebra \cite{Bourgine2021}. This approach should meet with the earlier attempt of Foda and Wheeler to build a B-type topological vertex \cite{FW,Foda2008}. We hope to be able to report on this problem soon.

Finally, we have been working here with the A-model formulation of topological strings. In the B-model, the connection with the KP hierarchy can also be seen using the Hermitian matrix model. In this context, the refinement is well-understood as the matrix model is replaced by a beta-ensemble \cite{Dijkgraaf2009}.\footnote{Note that the fermionic description is lost for both A and B models after the refinement.} It would be instructive to reproduce our derivation on the other side of the mirror.

\section*{Acknowledgements}
The author would like to thank Sasha Alexandrov, Yutaka Matsuo, Elli Pomoni and Kanehisa Takasaki for very helpful discussions. He is also very grateful to Pr. Kimyeong Lee and the Korea Institute for Advanced Study (KIAS) for their generous support in these difficult times. This research was partly supported by the Basic Science Research Program through the National Research Foundation of Korea (NRF) funded by the Ministry of Education through the Center for Quantum Spacetime (CQUeST) of Sogang University (NRF-2020R1A6A1A03047877).

\appendix

\section{Proofs}\label{AppA}
\subsection{Reminder on Miki's automorphism}
This reminder is a brief summary of the appendix A in \cite{Bourgine2018a} that is itself based on the original paper \cite{Miki2007}. We denote by $y^\pm(z)=\CS\cdot x^\pm(z)$ and $\xi^\pm(z)=\CS\cdot\psi^\pm(z)$ the image of the Drinfeld currents under Miki's automorphism. Just like the original currents in \ref{com_ak}, the S-dual currents can be decomposed in terms of modes 
\begin{equation}\label{def_y_xi}
y^\pm(z)=\sum_{k\in\mathbb{Z}}z^{-k}y_k^\pm,\quad \xi^\pm(z)=\sum_{k\geq0}z^{\mp k}\xi_{\pm k}^\pm=\xi_0^\pm\exp\left(\pm\sum_{k>0}z^{\mp k}b_{\pm k}\right),
\end{equation}
with $y_k^\pm=\CS\cdot x_k^\pm$, $b_k=\CS\cdot a_k$ and $\xi^\pm_{\pm k}=\CS\cdot\psi^\pm_{\pm k}$. Since $\CS$ is an automorphism, these new modes satisfy the same commutation relations as the original algebra, e.g.
\begin{align}
\begin{split}\label{com_bk}
&[b_k,b_l]=-(\g^{k\bc}-\g^{-k\bc})c_k\d_{k+l},\quad [b_k,y_l^\pm]=\pm\g^{\pm |k|\bc/2}c_k y_{l+k}^\pm,\\
&[y^+_k,y^-_l]=\left\{
\begin{array}{l}
\k\g^{-(k-l)\bc/2}\xi^+_{k+l},\quad k+l>0\\
\k\g^{-(k-l)\bc/2}\xi_0^+-\k\g^{(k-l)\bc/2}\xi_0^-,\quad k+l=0\\
-\k\g^{(k-l)\bc/2}\xi^-_{k+l},\quad k+l<0.\\
\end{array}
\right.
\end{split}
\end{align}

The expression for the S-dual modes $y_k^\pm$, $\xi_{\pm k}^\pm$ in terms of the original ones has been obtained by Miki in \cite{Miki2007},
\begin{align}
\begin{split}\label{Miki}
&y^\pm_k=(\pm)^k\g^{-(c\pm k\bc)/2}\s_1^{-(k-1)}\left(\text{ad}_{x_0^+}\right)^{k-1} x_{\mp1}^+,\quad y^\pm_{-k}=-(\pm)^k \g^{(c\mp k\bc)/2} \s_1^{-(k-1)}\left(\text{ad}_{x_0^-}\right)^{k-1} x^-_{\mp1},\\
&\xi_{\pm k}^\pm=-(\mp)^k(\g-\g^{-1})\s_1^{-(k-1)}\g^{\mp c}\text{ad}_{x_{\mp1}^\pm}\left(\text{ad}_{x_0^\pm}\right)^{k-2}x_{\pm1}^{\pm},\quad \xi_{\pm1}^\pm=\pm\g^{\mp c}(\g-\g^{-1})x_0^\pm,\quad \xi_0^\pm=\g^{\mp c}.
\end{split}
\end{align}
In these formulas, we denoted the adjoint action $\ad_A B=[A,B]$ and $\s_1=(q_1^{1/2}-q_1^{-1/2})(q_2^{1/2}-q_2^{-1/2})$.

\subsection{Proof of the refined exchange relations}
In order to prove the first exchange relation \ref{rel_xchg_ref}, we use the following fact: since $\rho^{(1,0)}$ is a representation and $\CT$ an automorphism, if two elements satisfy the exchange relation, so does their sum, product, commutator,... We have already shown that $b_{-1}\propto x_0^-$ obeys the exchange relation. In the same way, it is possible to show that the AFS relation \ref{AFS_lemma} with $e=x_k^-$ produces an exchange relation of the type \ref{rel_xchg_ref} where $b_{-k}$ is replaced by $x_k^-$. It follows from the coproduct \ref{coproduct} and the fact that the modes $x_k^-$ annihilate the vacuum in the vertical representation of levels $(0,1)$. We also notice that $c$ obeys the exchange relation but $\bc$ does not since $\CT\cdot\bc =c+\bc$. Since the expression \ref{Miki} for $\xi_{-k}^-$ involves only sums and products of $x_k^-$ and $c$, this operator obeys the exchange relation and so does $b_{-k}$ for $k>0$ after expansion on $z$. Note, however, that $b_1\propto x_0^+$ does not satisfy the exchange relation, and neither does the modes $b_k$ for $k>0$. The second exchange relation holds because the action of $\CT^{-1}$ simply shifts the modes $x_k^-\to x_{k+1}^-$ in the expression of $\xi_{-k}^-$ and thus the previous arguments hold as well in this case.

\subsection{Proof of the algebraic properties for the refined framing operator}
We present here the proof of the first identity in \ref{id_T_bk} for $k>0$. The other cases, namely $k<0$ and the second identity, can be proven using similar arguments. Furthermore, to avoid writing overcomplicated expressions, and since we work only in the module $\CF_0$ acted on by the representation $\rho_u^{(1,0)}$, we omit to indicate the representation of the quantum toroidal modes. We start with the case $k=1$. Once combined with Miki's transformation \ref{Miki}, the mode expansion \ref{def_y_xi} for $\xi^-(z)$ gives at first orders in $z$, 
\begin{equation}
b_{-1}=(\g-\g^{-1})x_0^-\implies \CT\cdot b_{-1}=(\g-\g^{-1})x_1^-.
\end{equation} 
Using the algebraic relations \ref{com_ak}, this can be written further
\begin{equation}
 \CT\cdot b_{-1}=-\g^{-c/2}c_1^{-1}(\g-\g^{-1})[a_1,x_0^-].
\end{equation} 
The action of the r.h.s. on the Macdonald states can be computed explicitly using the Pieri rules \ref{Pieri} for $a_1$ and the fact that $x_0^-\propto b_{-1}$ is diagonal (see \ref{action_bk}),
\begin{equation}
a_1\ket{P_\l}=\sum_{\sAbox\in R(\l)}r_\l^{(-)}(\Abox)\ket{P_{\l-\sAbox}},\quad x_0^-\ket{P_\l}=\dfrac{\Phi_1(\l,u)}{\g-\g^{-1}}\ket{P_\l}.
\end{equation} 
We do not need the explicit expression for the coefficients $r_\l^{(-)}(\Abox)$. Then, denoting $\chi_{\sAbox}=q_1^{i-1}q_2^{j-1}$ the content of a box $\Abox=(i,j)\in\l$, we compute
\begin{align}
\begin{split}
\CT\cdot b_{-1}\ket{P_\l}&=-\g^{-1/2}c_1^{-1}\sum_{\sAbox\in R(\l)}r_\l^{(-)}(\Abox)\left(\Phi_1(\l,u)-\Phi_1(\l-\Abox,u)\right)\ket{P_{\l-\sAbox}}\\
&=u^{-1}\g^{1/2}\sum_{\sAbox\in R(\l)}r_\l^{(-)}(\Abox)\chi_{\sAbox}^{-1}\ket{P_{\l-\sAbox}}\\
&=u^{-1}\g^{1/2}Fa_1 F^{-1}\ket{P_\l}.
\end{split}
\end{align}
Thus, we have shown the identity \ref{id_T_bk} for $k=1$. To simplify the upcoming formulas, we introduce the rescaled mode $\a_1=a_1/(\g-\g^{-1})$, so the previous identity writes $x_1^-=u^{-1}\g^{1/2}F\a_1 F^{-1}$.

Before addressing the general case, we would like to start with a short remark. Using the algebraic relations \ref{com_ak}, it is possible to write down
\begin{equation}
\psi_k^+=\k^{-1}\g^{kc/2}[x_0^+,x_k^-],\quad x_k^-=(-1)^k\g^{-kc/2}c_1^{-k}(\ad_{a_1})^k x_0^-.
\end{equation} 
Combining both, we arrive at an expression that be will useful later,
\begin{equation}\label{expr_psik}
\psi_k^+=\k^{-1}(-1)^kc_1^{-k}\ad_{x_0^+}(\ad_{a_1})^k x_0^-.
\end{equation} 

In a similar way, we write the modes $\xi_{-k}^-=-\k^{-1}\g^{(k+2)\bc/2}[y_{-(k+1)}^+,y_1^-]$ for $k>1$ as a commutator of the modes $y_k^\pm$ using the relations \ref{com_bk}, and then use Miki's transformation \ref{Miki} to write down
\begin{equation}
y_{-(k+1)}^+=-\g^{c/2}\g^{-(k+1)\bc/2}\s_1^{-k}(\ad_{x_0^-})^kx_{-1}^-,\quad y_1^-=-\g^{-(c-\bc)/2}x_1^+.
\end{equation} 
As a result, we find
\begin{equation}
\xi_{-k}^-=\k^{-1}\g^\bc\s_1^{-k}\ad_{x_1^+}(\ad_{x_0^-})^kx_{-1}^- \implies \CT\cdot \xi_{-k}^-=\k^{-1}\g^{c+\bc}\s_1^{-k}\ad_{x_0^+}(\ad_{x_{1}^-})^kx_{0}^-.
\end{equation} 
Using the identity found previously for $x_1^-$, and the fact that $x_0^-$ commutes with $F$ (they are both diagonal in the Macdonald basis), we have
\begin{equation}
e^{zu \ad_{x_1^-}}x_0^-=e^{zux_1^-}x_0^-e^{-zux_1^-}=Fe^{z\g^{1/2}\a_1}x_0^-e^{-z\g^{1/2}\a_1}F^{-1}=Fe^{\ad_{z\g^{1/2}\a_1}}x_0^-F^{-1}.
\end{equation} 
Expanding in powers of $z$, we deduce that
\begin{equation}
\CT\cdot \xi_{-k}^-=\k^{-1}\g^{1+k/2}\s_1^{-k}u^{-k}\ad_{x_0^+}\left(F\left((\ad_{\a_1})^kx_{0}^-\right)F^{-1}\right),
\end{equation} 
where we have also identified the central charges $(c,\bc)$ with the levels $(1,0)$. Then, we use the general result $[A,FBF^{-1}]=F[F^{-1}AF,B]F^{-1}$, together with the fact that $x_0^+$ also commutes with $F$, to write down
\begin{equation}
\CT\cdot \xi_{-k}^-=\k^{-1}\g^{1+k/2}\s_1^{-k}u^{-k}F\left(\ad_{x_0^+}(\ad_{\a_1})^kx_{0}^-\right)F^{-1}.
\end{equation} 
Comparing with the formula \ref{expr_psik} for the modes $\psi_k^+$, we find that
\begin{equation}
\CT\cdot \xi_{-k}^-=\g^{1+k/2}(-u)^{-k}\ F\psi_k^+F^{-1}\implies \CT\cdot\xi^-(z)=\g F\psi^+(-\g^{-1/2}uz^{-1})F^{-1}.
\end{equation} 
The first identity in \ref{id_T_bk} with $k>0$ follows from the expansion of the exponentials, while the difference of zero modes takes care of the extra factor $\g$. Applying the same method to $\CT\cdot\xi^+(z)$, we can prove the following identities that produce the other cases of the relations \ref{id_T_bk} by expansion,
\begin{align}
\begin{split}
&\CT\cdot\xi^+(z)=\g^{-1} F\psi^-(-\g^{-1/2}uz^{-1})F^{-1},\\
&\CT^{-1}\cdot\xi^-(z)=\g F^{-1}\psi^-(-\g^{1/2}u^{-1}z)F,\\
&\CT^{-1}\cdot\xi^+(z)=\g^{-1} F^{-1}\psi^+(-\g^{1/2}u^{-1}z)F.
\end{split}
\end{align}

\bibliographystyle{../../utphys}
\bibliography{../Real_TV_IH}
\end{document}